\documentstyle[psfig]{mn}
\def\H0{{\it H}$_0$}
\def\Ms{{\it M}$_\odot$}
\def\Ls{{\it L}$_\odot$}
\def\q0{{\it q}$_0$}
\def\kmps{km~s$^{-1}$}

\def\ergps{erg~s$^{-1}$}
\def\kmpspMpc{km~s$^{-1}$~Mpc$^{-1}$}
\def\Ms{{\it M}$_\odot$}
\def\micron{$\mu$m}

\def\nH{$N_{\rm H}$\thinspace} 
\def\psqcm{cm$^{-2}$}
\def\ergpspsqcm{erg~cm$^{-2}$~s$^{-1}$}

\def\ni{\noindent}

\def\Zs{$Z_{\odot}$}

\def\cps{ct\thinspace s$^{-1}$}

\def\Av{$A_{\rm V}$\/}
\def\pcubcm{cm$^{-3}$}

\title[X-ray emission from powerful FIRGs]
{An X-ray investigation of powerful far-infrared galaxies}
\author[K. Iwasawa]
{\parbox[]{6.5in}{K. Iwasawa}\\
\\
Institute of Astronomy, Madingley Road, Cambridge CB3 0HA\\}

\date{}

\begin{document}

\maketitle

\begin{abstract}
We present ASCA results on four prototype powerful far-infrared galaxies,
Mrk231, Mrk273, Arp220 and NGC6240. The soft X-ray spectra show signatures
of thermal emission with temperatures of (0.5--1)$\times 10^7$ K, which is 
probably produced in starbursts. Their soft X-ray (0.5--2 keV) 
luminosities range from $4\times 10^{40}$\ergps\ to $7\times 10^{41}$\ergps.
The X-ray properties are examined in the context of a starburst.
Evidence for a heavily obscured active nucleus is found in Mrk273 and 
NGC6240. The ASCA spectra of both galaxies show strong iron K emission line features.
The hard X-ray emission ($> 3$ keV) in NGC6240 is most likely
the reflected light of a hidden QSO whose intrinsic luminosity
is suspected to be $\sim 10^{45}$\ergps, comparable to the far-IR luminosity.
The 2--10 keV emission, possibly related to an AGN, is found in Mrk231.
The observed 2--10 keV luminosity is 
only $\sim 2\times 10^{42}$\ergps, and the origin of the hard X-ray emission
is uncertain due to 
the low quality of the present data. No evidence for an AGN is found 
in Arp220 in the X-ray data. However, the soft X-ray emission originating
in a starburst is also as weak as the H$\alpha$ nebula and near-IR continuum
despite the large far-IR excess. The possible existence of a powerful but
heavily obscured (Compton-thick) AGN is discussed.
\end{abstract}


\section{introduction}

The energy source of a class of powerful far-infrared galaxies (FIRGs), 
which emit the bulk of their bolometric luminosity in the far-IR domain
(Soifer et al 1984; Sanders \& Mirabel 1996 and references therein), 
is still a major issue under debate.
The large luminosity ($> 10^{11}$\Ls) of the IR emission, which is
likely due to dust reradiation, is comparable with that of QSOs.
The far-IR energy distribution of the prototype FIRGs, NGC6240, Mrk273 and Arp220, 
is similar to starburst galaxies like M82, unlike the `warm' IRAS colour 
(e.g., $S_{25}/S_{60}\geq 0.2$) typical of Seyfert galaxies 
and QSOs (DeGrijp et al 1992).

The optical spectra of the nuclear regions of FIRGs resemble
LINER or Seyfert 2s (Sanders et al 1988), but they appear to be weak
in X-ray. 
A search for X-ray sources associated with powerful FIRGs
in the HEAO-1 A1 all sky survey data failed to detect any (Rieke 1988).
No detectable hard X-rays in the 50--700 keV band for Arp220,
Mrk273 and Mrk231 with the Compton Gamma-Ray Observatory (CGRO)--OSSE 
(Dermer et al 1997) rules out
luminous Compton-thin X-ray sources (\nH $<1\times 10^{24}$\psqcm).
No detection of strong hard X-ray emission means that X-rays from a central
source are attenuated by a heavy obscuration, if AGNs power FIRGs.

On the other hand, various observations suggest that starburts play
a significant role. 
Direct evidence for a massive star population has been provided by
the detection of near-IR CO bandhead absorption at 2.3\micron,
which is produced in the stellar atmospheres of red giants and supergiants
(Rieke et al 1980, 1985; Walker Lebofsky \& Rieke 1988; Lester et al 
1990; Doyon et al 1994; Ridgway, Wynn-Williams \& Becklin et al 1994).

A large (tens kpc scale) extended optical emission-line nebula
is often found around FIRGs (Heckman, Armus \& Miley 1987;
Armus, Heckman \& Miley 1990).
Studies of the gas kinematics in the nebulae (e.g., split line-profiles)
have revealed galactic-scale outflow (`superwind', Heckman, Armus \& 
Miley 1990) with velocities of a few hundreds--1000 \kmps.
A collection of supernovae (SNe) and stellar winds from massive stars
in a nuclear starburst can drive this outflow.
The LINER-type excitation of the nebular emission-line gas is consistent 
with shock-heating by the superwind, which is also expected to
produce soft X-rays (Chevalier \& Clegg 1985; Tomisaka \& Ikeuchi 1988).
Some FIRGs show extended soft X-ray emission observed with Einstein 
Observatory and ROSAT (Fabbiano 1988; Strickland, Ponman \& Stevens 1997; 
Heckman et al 1996; Dahlem et al 1997; Schulz et al 1998; Komossa,
Schulz \& Greiner 1998).
Since this process of generating X-rays is thought to be not very efficient 
(log($L_{\rm X}/L_{\rm FIR})\sim 10^{-4}$), the X-ray quiet nature of 
FIRGs would not be surprising.

The weak reddening-corrected luminosity (and small equivalent width) 
of Br$\alpha $ and Br$\gamma $ in some FIRGs (e.g., Arp220) 
have raised the `ionization photon deficit' problem with a starburst model:
ionizing photons predicted from the bolometric luminosity 
significantly overproduce the near-IR lines observed (DePoy,
Becklin \& Geballe 1987; Goldader et al 1995; Armus et al 1995;
Leitherer \& Heckman 1995).
Although this has sometimes been claimed as evidence against 
a starburst model, it essentially means that the major energy source 
(whether starburst or AGN) is heavily obscured in dust even in 
the near-IR band.
The mid-IR line spectroscopy of powerful FIRGs with the Infrared Space
Observatory (ISO) Short Wavelength Spectrometer (SWS) found such
a large extinction (\Av $\sim $40--50) for starburst regions (Lutz et al 1996;
Strum et al 1996; Genzel et al 1998), and therefore 
they favour young starbursts as the primary energy source in FIRGs.

However, the existence of an even more deeply buried AGN is still viable.
The intensive starburst in the outer part of an obscuring region may
mask AGN acitivity.
The nearby FIRG, NGC4945, for which the mid-IR properties 
are starburst-like, harbours a variable X-ray source absorbed by \nH\
$\approx 5\times 10^{24}$\psqcm\ (Iwasawa et al 1993; Done, Madejski
\& Smith 1996), implying a heavily obscured Seyfert nucleus.
The ASCA data on the prototype powerful FIRG, NGC6240, suggests 
the presence of a QSO nucleus deeply buried in obscuration (Mitsuda 1995; 
Kii et al 1997; Iwasawa \& Comastri 1998 and see below).

High sensitivity observations in hard X-ray band provides a
unique probe to such heavily obscured AGNs.
The upper-limits of hard X-ray emission for powerful FIRGs obtained from
previous observations (Rieke 1988; Eales \& Arnoud 1988; Dermer et al 
1997) are still consistent with the hypothesis that the FIRGs contains
Compton-thick sources like NGC1068:
this classical Seyfert 2 galaxy is believed to harbour a
powerful Seyfert 1 nucleus behind an obscuring torus (e.g., Antonucci \&
Miller 1985; Miller, Goodrich \& Mathews 1991).
The observed hard X-rays are only due to 
weak reflected light of the central source, 
whose direct radiation is completely blocked by an extremely thick absorber
(\nH $\gg 10^{25}$\psqcm, Koyama et al 1989; Matt et al 1997)
in the line of sight. It should be noted that 
the hard X-ray flux level is three orders of magnitude below the far-IR emission,
which implies the detection limits of HEAO-1 and OSSE are insufficient
for distant FIRGs.
This classic example makes a sensitive X-ray survey of the FIRGs 
searching for AGNs still worthwhile. 

The Compton-thick situation can be more likely in FIRGs than 
normal Seyfert 2 galaxies.
The large amount of molecular gas concentrated in the central region 
(sub-kilo parsec in size) of powerful FIRGs (Sanders, Scoville \& Soifer 1991; 
Scoville et al 1991; Solomon, Downes \& Radford 1992) 
and its high density ($\sim 10^4-10^5$ \pcubcm, Solomon et al 1992; Bryant \&
Scoville 1996; Scoville, Yun \& Bryant 1997) would easily make up 
a large column density (e.g., $> 10^{24}$\psqcm) to suppress the direct
X-ray radiation from a central source.

Even reflected light would be difficult to detect 
in the soft X-ray band as observed with ROSAT, 
since moderate absorption 
($\sim 10^{22}$\psqcm) would be expected because of 
the high extinction for the nuclear regions, and it suppresses
the soft X-ray light. 
Therefore X-ray observations above $\sim $2 keV are important to
observe AGN free from absorption.

The reflection-dominated X-ray spectrum is characterized by a flat hard 
X-ray continuum and a strong iron 
K line at 6.4 keV, as seen in NGC1068 (Marshall et al 1993;
Ueno et al 1994; Iwasawa, Fabian \& Matt 1997).
Therefore X-ray spectra are also important to assess the origin of 
observed X-ray emission.

Using good spectral resolution and high sensitivity to X-rays above 2 keV,
where AGN emission would dominate, we study four prototype powerful
FIRGs, Mrk231, Mrk273, Arp220 and NGC6240.
We discuss the implications on starbursts and AGNs from the X-ray data, and
the primary energy sources of the FIRGs.

\section {Selected sample}

We present results on four prototype powerful FIRGs: Arp220, NGC6240,
Mrk231 and Mrk273 (Table 1). 
They are all among the infrared colour-selected sample (ICSS) galaxies 
(Armus, Heckman \& Miley 1990)
 and lie in a well-defined part of 
IRAS colour-colour diagram
($\alpha_{25-60}<-1.5$, $\alpha_{60-100}>-0.5$)
which is distinctive from where other class of objects occur:
their `tepid' characteristic temperatures are different from
that of normal galaxies or Seyferts/QSO.
The four galaxies are good representative of the particular class
of objects in terms of large IR luminosity ($\sim 10^{12}$\Ls), 
characteristic far-IR spectrum, distorted host galaxy,
and concentrated large molecular gas content ($M_{\rm H_2}\geq 10^9$\Ms,
Sanders, Scoville \& Soifer 1991; Scoville et al 1991).
A double nucleus is found in all galaxies except Mrk231.
The optical spectra of their nuclear regions are considerably red
and the emission-line properties are classified as 
LINER (NGC6240) Seyfert-2 (Mrk273 and Arp220) and Seyfert-1 (Mrk231).
Kinematic signatures of galactic-scale outflow (e.g., split line profiles,
Heckman et al 1990; Hamilton \& Keel 1987) have been found in 
the circumnuclear regions, and large extended H$\alpha $ nebulae (tens 
kpc) were imaged for all the galaxies by Armus et al (1990).
Weak soft X-ray emission has been detected from all the galaxies with
Einstein Observatory and ROSAT. No detection of X-rays above $\sim $3 keV has
been reported for them before the ASCA observations.

\section{observations and data analysis}

\begin{table*}
\begin{center}
\caption{The observed sample. The distances to the galaxies are calculated,
assuming \H0\ = 75 \kmpspMpc. The infrared luminosity, $L_{\rm IR}$, for 
8--1000 \micron\ is taken from Soifer et al (1987), which is based on the 
IRAS fluxes. $S_{60}/S_{25}$ is the ratio of flux densities between 60\micron\
and 25\micron, measured by IRAS.}
\begin{tabular}{lcccccc}
Galaxy & IRAS number & $z$ & Distance & log ($L_{\rm IR}$/\Ls) & 
$L_{\rm FIR}/L_{\rm B}$ & $S_{60}/S_{25}$ \\
& & & Mpc & & & \\
Mrk 231 & 12540+5708 & 0.042 & 174 & 12.50 & 20 & 4.6 \\
Mrk 273 & 13428+5608 & 0.038 & 152 & 12.10 & 29 & 3.7 \\
Arp 220 & 15327+2340 & 0.018 & 73 & 12.11 & 150 & 13.1 \\
NGC 6240 & 16504+5235 & 0.024 & 100 & 11.77 & 45 & 6.6 \\
\end{tabular}
\end{center}
\end{table*}

\begin{table*}
\begin{center}
\caption{ASCA observations. $^{\ast}$ This is the 
S1 exposure time while 12 ks for S0. The count rates are corrected for
background.}
\begin{tabular}{lcccc}
Galaxy & Date & SIS mode & Exposure (SIS/GIS) & Count rate (S0/G2) \\
& & & $10^3$ s & $10^{-2}$ \cps \\
Mrk 231 & 1994 Dec 05 & Faint+Bright/2CCD & 25/21 & 1.0/0.7 \\
Mrk 273 & 1994 Dec 27 & Faint/1CCD & 46/41 & 1.2/0.7 \\
Arp 220 & 1993 Jul 26 & Faint/1CCD & 26$^{\ast}$/32 & 1.0/0.5 \\
NGC 6240 & 1994 Mar 27 & Faint/2CCD & 39/41 & 4.9/2.8 \\
\end{tabular}
\end{center}
\end{table*}

ASCA observations of the sample FIRGs are summarized in Table 2.
The data were retrieved from the ASCA archive database maintained by
ASCA Guest Observaer Facility (GOF) at Goddard Space Flight Center.
Standard calibration and data reduction technique (FTOOLS)
are used. Spectral analysis was performed using
XSPEC.


ASCA carries four X-ray telescopes (XRT)
with focal plane detectors for each (Tanaka, Inoue \& Holt 1994).
At the moderate spatial resolution of the ASCA XRT 
(a half power diameter $\sim $3 arcmin, Serlemitsos et al 1995)
extended X-ray emission from the sample galaxies reported by ROSAT PSPC/HRI
observations cannot be resolved.

There are two types of focal plane detectors (a brief description is 
in Tanaka et al 1994).
The Solid state Imaging Spectrometer (SIS; S0 and S1) is 
an X-ray sensitive CCD detector (0.4--10 keV band) with higher spectral 
and spatial resolutions than the Gas Imaging Spectrometers (GIS; G2 and
G3 for the 0.7--10 keV band). 
The higher sensitivity of the SIS to soft X-rays ($< 2$ keV), combined
with the spectral resolution, is suitable to study thermal emission
with a temperature of $\sim 10^7$ K, often associated to starburst
while the GIS has a larger effective area above 5 keV which is
useful to detect hard X-ray emission originating in AGN.

The thermal emission model for optically-thin, collisional ionization
equilibrium plasma, 
MEKAL, is used in the spectral fits,
with solar abundances by Feldman (1992).
The absorption cross sections are taken from Morrison \& McCammon (1983).
The column densities of Galactic absorption for each galaxy are 
derived from the HI measurements by Dickey \& Lockman (1990).
Quoted errors to best-fit spectral parameters are 90 per cent confidence
regions for one parameter of interest.
The Hubble constant of $H_0 = 75$ \kmpspMpc is used 
in deriving distances of galaxies (Table 1).

\section{Results}

\subsection{Mrk231}

This galaxy has a single nucleus, although 
a faint double tidal-tail in the visual image suggests that this system 
could be a recently completed merger (Huchings \& Neff 1987;
Sanders et al 1988). The second nucleus claimed by Armus et al (1994) appears
to be star-forming knots resolved in the HST image (Surace et al 1998).
Besides its large infrared luminosity, Mrk231 has many other outstanding
properties.
AGN-like phenomena dominate the radio, optical and near-IR band, but 
the nature appears to be unusual.
The optical spectrum of the nucleus shows broad (FWHM$\approx 4200$\kmps)
Balmer lines, strong FeII emission and several absorption line systems
blueshifted by up to 8240 \kmps (Boksenberg et al 1977).
The multiple transitions of FeII emission dominate the optical spectrum while 
there is little evidence for narrow emission lines.
The nonthermal radio continuum of Mrk231 is variable on time scales of 
months or years (e.g., Condon, Frayer \& Broderick 1991).

The optical continuum of the Seyfert 1 nucleus is considerably red 
(Weedman 1973; Boksenberg et al 1977).
This trend continues to the UV band 
but with a blue continuum upturn shortward of 
2400\AA ~probably due to a hot star cluster (Smith et al 1995).
Unusually high, wavelength dependent optical/UV polarization (10--20 per cent 
at shorter wavelengths than 4000\AA, Thompson et al 1980; Schmidt \& 
Miller 1985; Smith et al 1995) is likely to arise in dust scattering
with dilution by the starlight observed in UV.
The polarization degree changing across the broad H$\alpha$ profile
indicates that some of the broad component is dust-scattered 
(Smith et al 1995).
Various observations suggest that the active nucleus is a BAL QSO,
which is mostly
obscured by dust and low-ionization gas (e.g., Hamilton \& Keel 1987;
Smith et al 1995), with strong FeII emission.

However, the type-1 AGN characteristics disappear 
in the mid-IR and X-ray bands.
No high excitation lines were detected in the ISO/SWS spectrum
(Rigopoulou et al 1998).
Hard X-ray emission is also very faint ($<6.5\times 10^{-12}$\ergpspsqcm,
Rieke 1988).
Since the nuclear absorption (\nH $\sim 4\times 10^{21}$\psqcm)
implied from the optical reddening (\Av\ $\sim 2$, Boksenberg et al 1977)
should be no longer opaque to X-rays of energies above 2 keV,
the lack of detection of strong hard X-rays is puzzling
for Seyfert-1/QSO nuclei.

It is interesting to note that estimates of reddening towards the mid-IR
nucleus are much larger than that deduced for the optical continuum:
\Av\ $\sim 20$ from the silicate absorption feature at 10\micron 
(Roche, Aitken \& Whitmore 1983);
and \Av\ $\sim 40$ from the [SIII] line ratio (18.7\micron\ and 
33.5\micron, Rigopoulou et al 1998).

Weak soft X-rays have been detected with the ROSAT PSPC.
Although Rigopoulou, Lawrence \& Rowan-Robinson (1996), Rush et al (1996) 
and Wang, Brinkmann \& Bergeron (1996) fitted the
PSPC spectrum with a power-law or bremsstrahlung models,
we have found that thermal emission from optically thin
plasma, expected from a starburst, is a likely explanation 
(see below and Table 3).
At a temperature around $10^7$K, an X-ray spectrum is of an emission-line
dominated regime so a bremsstrahlung model is not appropriate for it.
The evidence for thermal X-ray emission 
is confirmed by the ASCA data.
Above the ROSAT band, we detect hard X-ray emission in the ASCA data.

\subsubsection{The ROSAT PSPC data}


\begin{table}
\begin{center}
\caption{Spectral fits to the ROSAT PSPC data on Mrk231. Galactic absorption,
\nH $= 1.26\times 10^{20}$\psqcm ~is assumed for both model. (1) Photon-index
of power-law; (2) excess absorption above the Galactic value; (4) temperature
of MEKAL thermal emission spectrum; (5) metal abundance.}
\begin{tabular}{ccc}
\multicolumn{3}{c}{Power-law} \\
(1) & (2) & (3) \\
$\Gamma$ & \nH & $\chi^2$/dof \\
& $10^{20}$\psqcm & \\
$1.73\pm 0.15$ & --- & 36.82/21 \\
$2.22\pm 0.48$ & $2.4\pm 2.0$ & 33.22/20 \\[5pt]
\multicolumn{3}{c}{Thermal emission} \\
(4) & (5) & (6) \\
$kT$ & $Z$ & $\chi^2$/dof \\
keV & \Zs & \\
$0.88^{+0.27}_{-0.17}$ & $0.10^{+0.08}_{-0.04}$ & 19.68/20 \\
\end{tabular}
\end{center}
\end{table}

Mrk231 was observed with the ROSAT PSPC on 1991 June 7--8.
The energy spectrum of the 24 ks ROSAT PSPC data is reanalysed.
Fitting with an absorbed power-law model to the 0.2--2 keV data
gives a photon-index of $2.22\pm 0.48$ plus intrisic
absorption of \nH = $(2.4\pm 2.0)\times 10^{20}$\psqcm, which are 
consistent with results by Rigopoulou et al (1996).
This fit leaves significant residuals around 0.9 keV, which is
possibly an emission-line bump due to Fe L.
The thermal emission model gives a much better fit
(see Table 3) with a temperature of
$kT = 0.88^{+0.27}_{-0.17}$ keV and abundance, $Z=0.10^{+0.08}_{-0.04}$\Zs.
No significant absorption excess above Galactic value is required.
The 0.1--2 keV flux estimated from the thermal model is
$1.27\times 10^{-13}$\ergpspsqcm.

\subsubsection{The ASCA data}


\begin{figure}
\centerline{\psfig{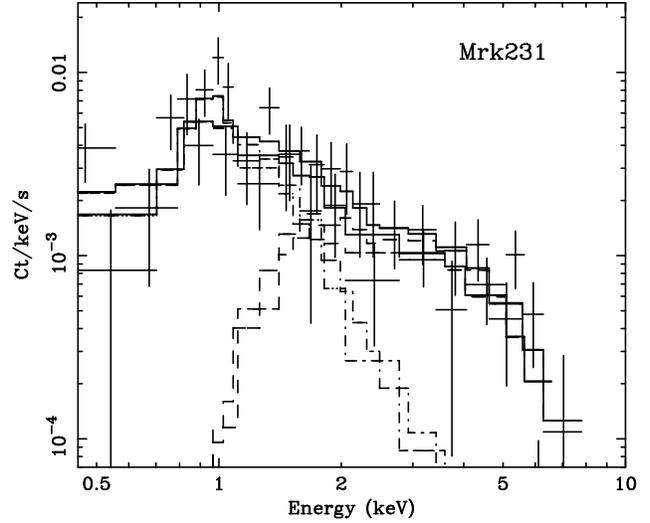}}
\caption{The ASCA SIS spectrum of Mrk231. The data were fitted by 
a thermal spectrum plus an absorbed power-law. Best-fit parameters are 
given in Table 4.}
\end{figure}


\begin{table*}
\begin{center}
\caption{Spectral fits to the ASCA data on Mrk231 and Arp220. 
Spectral fittings were performed to the four detectors jointly 
with a model consisting a thermal emission 
spectrum and an absorbed power-law. No extra absorption above Galactic
extinction (\nH = $1.3\times 10^{20}$\psqcm ~for Mrk231 and 
\nH = $ 3.7\times 10^{20}$\psqcm ~for Arp220) is assumed for the thermal 
emission component. $\ast$: normalization of the thermal emission model
in [$10^{-18}/(4\pi D^2)]\int n^2dV$, where $D$ is the distance of the source
and $n$ is electron density in cm$^{-3}$. $\dag$: Normalization of power-law
at 1 keV in unit of $10^{-5}$photons keV$^{-1}$cm$^{-1}$s$^{-1}$. 
Note that the Arp220 data include the southern source (see text).}
\begin{tabular}{lccccccc}
Galaxy & \multicolumn{3}{c}{Thermal emission} & \multicolumn{3}{c}{Power-law}
& $\chi^2$/dof \\
& $kT$ & Z & $A_{\rm th}$ & $\Gamma$ & $A_{\rm PL}$ & \nH & \\
& keV & \Zs & $\ast$ & & $\dag$ & $10^{22}$\psqcm & \\[5pt]
Mrk 231 & $0.98_{-0.09}^{+0.51}$ & $0.08^{+0.2}_{-0.05}$ & 2.8 & 
$1.2^{+0.7}_{-0.5}$ & 6.6 & $2.1^{+2.5}_{-1.5}$ & 98.07/116 \\
& $0.98^{+0.66}_{-0.25}$ & $0.06^{+0.2}_{-0.04}$ & 3.2 & 
1.8 & 18 & $3.9^{+3.6}_{-0.9}$ & 100.2/117 \\[5pt]
Arp220 & $0.76^{+0.11}_{-0.11}$ & 0.31($\geq 0.1$) & 
0.8 & 1.8 & 0.4 & 0.8($<2.7$) &
69.72/69 \\
\end{tabular}
\end{center}
\end{table*}

The signal-to-noise of the ASCA data is rather poor because of the short 
exposure time for this faint X-ray source.
As we showed above for the ROSAT PSPC data, 
the soft X-ray emission appears to be thermal emission.
The soft portion of the ASCA spectrum also indicates a possible Fe L emission
peak around 0.9 keV and is described well with the thermal emission model.

We find a hard-spectrum component as well as the thermal emission
in the ASCA data (see Fig 1). 
When the data above 2 keV from four detectors are fitted by a power-law 
modified only by Galactic absorption (\nH $=1.26\times 10^{20}$\psqcm),
a photon index of $0.8^{+0.4}_{-0.5}$ is obtained.
This very hard X-ray spectrum above 2 keV is unusual for Seyfert 1 
galaxies which generally have photon indices around 1.8 in the ASCA band
(e.g., Reynolds 1997).
If the hard-spectrum component originates in a central source of 
the Seyfert 1 nucleus, it could have been altered by
absorption and/or reflection.

The whole ASCA spectrum is modelled by a thermal emission model 
plus an absorbed power-law. 
The photon index of the power-law cannot be well constrained when 
it is left to be a free parameter (Table 4), 
$\Gamma = 1.8$, the mean
ASCA value for Seyfert 1 galaxies, is also consistent.
Results are summarized in Table 4.
The temperature and metallicity of the thermal emission model obtained 
are consistent with the ROSAT results, within the uncertainties.
The hard X-ray component can be approximated by an absorbed power-law
which has large absorption with a column density of
\nH$\simeq 4\times 10^{22}$\psqcm, when the $\Gamma = 1.8$ power-law
continuum is assumed. As shown above, a flat continuum 
($\Gamma\sim 0.8$) with smaller absorption is also acceptable.

The observed total fluxes in the 0.5--2 keV and 2--10 keV bands are
$1.2\times 10^{-13}$\ergpspsqcm ~and $6\times 10^{-13}$\ergpspsqcm,
respectively. Using the best-fit spectral model, the 0.5--2 keV 
luminosity of the thermal emission is $4\times 10^{41}$\ergps,
and the absorption corrected 2--10 keV luminosity of the power-law
is $2.2\times 10^{42}$\ergps.

The flat spectrum in the hard X-ray band ($>3$ keV) is also 
consistent with a reflection-dominated spectrum.
In this case, a strong iron K line feature would be observed on top of 
a very flat continuum, like
NGC1068 and NGC6240.
Unfortunately, the poor signal-to-noise of the present data can provide
only a loose constraint on an iron line strength;
the 90 per cent upper limit on EW for a narrow 
line at 6.4 keV is about 0.9 keV.

One might consider the possibility that the observed entire spectrum can be
explained by a power-law continuum of the
Seyfert nucleus modified 
by the multiple low-ionization absorption systems observed in
the optical band which could impose absorption edges due to 
partially ionized oxygen [OVII(0.74 keV) and OVIII(0.87 keV)]
onto the power-law continuum.
However, this possibility can be ruled out.
The inclusion of an absorption edge into a power-law to account for
the spectral break between 1 and 2 keV seen in the ASCA data however
gives a significantly worse fit to the thermal and power-law composite
model ($\Delta\chi^2\sim 11$). The obtained edge energy ($1.06\pm 0.05$
keV) is too high to match any blueshifted oxygen edges expected from
known absorption systems.

\subsection{Mrk273}


\begin{figure}
\centerline{\psfig{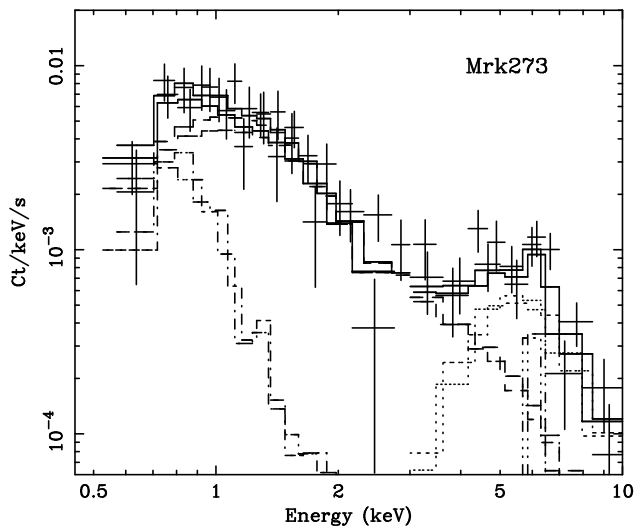}}
\caption{The ASCA spectrum of Mrk273. The data were fitted with 
a two-temperature thermal spectrum, absorbed power-law plus a gaussian line
for the Fe K.}
\end{figure}

Mrk273 shows a highly distorted optical morphology with a $\sim 1$ 
arcmin-long tail to the south and a multiple core 
(e.g., Sanders et al 1988). 
Radio and near-IR imagings show a double nucleus clearly (Condon et al 1991;
Majewski et al 1993; Knappen et al 1997). Note that the SE source resolved 
by Condon et al (1991) may be a background radio source while a weaker
SW source associated with the secondary near-IR/optical peak has been 
found (Knappen et al 1997).
The nuclear optical spectrum resembles that of 
a Seyfert 2 nucleus ([OIII]$\lambda 5007$/H$\beta
\sim 10$ and emission-line width of FWHM$\sim 700$\kmps)
with a heavy reddening implied by a large Balmer decriment 
(\Av $\sim 3$, Koski et al 1978).

\subsubsection{The X-ray companion source}

A weak soft X-ray source ($f_{\rm 0.1-2keV}\sim 1\times 10^{-13}$\ergpspsqcm)
has been detected with ROSAT (Turner, Urry \& Mushotzky 1993).
An X-ray companion source with comparable X-ray brightness was also 
found at 1.3 arcmin NE of Mrk273 (Turner et al 1993), 
which has been identified with a dwarf galaxy with Seyfert-2 emission-line
characteristics lying at the same redshift (Xia et al 1998).
In the 0.5--2 keV image of the SIS, the companion source appears to be
fainter (at most 40 per cent of Mrk273 in count rate).
According to the spectral studies of the ROSAT PSPC data (Turner et al 1993;
Xia et al 1998), the companion source has a spectrum similar to or
slightly harder than that of Mrk273. Therefore the relatively better efficiency
towards 2 keV of the SIS compared with the PSPC 
cannot be the explanation for the apparent weakness
of the companion source during the ASCA observation.
All the ROSAT PSPC and HRI observations of Mrk273 were carried out between
1992 May and June. The companion X-ray source has probably faded by factor 
of 2 or more in 2.5 yr.
There is a possible extent to the NW of Mrk273 which was not seen by 
the PSPC or HRI images.

The brightest source is found at the same position both in the
0.6--3 keV and 4--10 keV images.
The data used for spectral analysis were taken from a circular region
with a radius of 3 arcmin which also contains the dwarf galaxy.
The X-ray fluxes observed with the ASCA SIS are
$1.7\times 10^{-13}$\ergpspsqcm\ in the 0.5--2 keV band and 
$7.0\times 10^{-13}$\ergpspsqcm\ in the 2--10 keV band. 


\begin{figure}
\centerline{\psfig{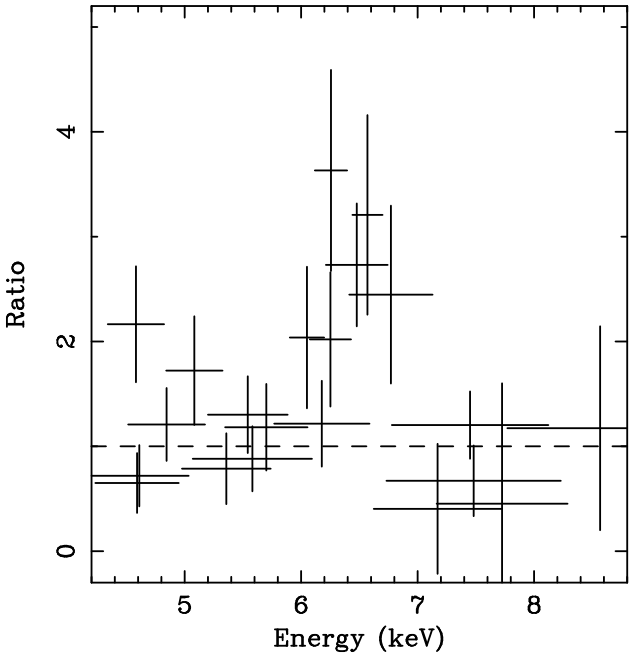}}
\caption{The iron K line feature of Mrk273. The ratio of the ASCA data
and the best-fit absorbed power-law continuum is plotted. The energy scale is 
corrected for the redshift ($z=0.038$). This feature is consistent with
a single narrow-line at 6.4 keV.}
\end{figure}

\subsubsection{The ASCA spectrum}


\begin{table*}
\begin{center}
\caption{Results of spectral fits to the ASCA data of Mrk273. The 0.5--10 keV
SIS data and the 1--10 keV GIS data are fitted jointly. The value of $\chi^2$
for this fit is 110.4 for 129 degrees of freedom. $A$ and $A_{\rm PL}$ 
are normalizations of the MEKAL and power-law (PL) models at 1 keV
in unit of $^{\ast } 10^{-14}(4\pi D^2)^{-1}\int dV$, 
and $^{\dag}$ ph\thinspace
keV$^{-1}$s$^{-1}$cm$^{-2}$, respctively. The PL$_1$ can be replaced by a
thermal emission model with a temperature of $\sim 5$ keV, without significant
difference in statistics. The EW of the line is $\sim 520$ eV.}
\begin{tabular}{lcccc}
 & $kT$ & $Z$ & $A$ & \nH \\
 & keV & \Zs & $\ast$ & \psqcm \\[5pt]
MEKAL & $0.47^{+0.24}_{-0.15}$ & $0.23(>0.03)$ & $8.9\times 10^{-5}$ & 
 $1.1\times 10^{20}$ \\[10pt]
& $\Gamma$ & $A_{\rm PL}$ & \nH & \\
& & $\dag $ & \psqcm & \\[5pt]
PL$_1$ & $1.7^{+1.0}_{-0.6}$ & $6.4\times 10^{-5}$ & 
   $1^{+1}_{-1}\times 10^{21}$ & \\[10pt]
PL$_2$ & 1.8 & $5.2\times 10^{-4}$ & 
   $4.3_{-1.3}^{+1.2}\times 10^{23}$ & \\[10pt]
 & \multicolumn{2}{c}{$E_{\rm line}$} & \multicolumn{2}{c}{$I_{\rm line}$} \\
 & \multicolumn{2}{c}{keV} & 
   \multicolumn{2}{c}{ph\thinspace s$^{-1}$cm$^{-2}$}\\[5pt]
Gaussian & \multicolumn{2}{c}{$6.39_{-0.20}^{+0.22}$} & 
   \multicolumn{2}{c}{$1.0_{-0.5}^{+0.9}\times 10^{-5}$} \\
\end{tabular}
\end{center}
\end{table*}

The ASCA spectrum of Mrk273 (Fig. 2) is complex, as seen in classical
Seyfert 2 galaxies.
Below 3 keV, the SIS spectrum can be roughly described with a power-law
with photon-index of $\sim $2.4 modified by absorption 
of \nH $\sim 1\times 10^{21}$\psqcm\ ($\chi^2 = 42.60$ for 53 degrees
of freedom). Although the data are noisy,
systematic line-like residuals around 0.7--0.8 keV remain after the
fit, which is probably attributed to Fe L emission.
If the Fe L emission originates in thermal emission, the temperature 
of the gas would be $kT \sim$0.4--0.5 keV, which is too low to explain
the observed spectrum up to 3 keV.
Therefore it is plausible that the 0.4--3 keV spectrum consists of 
low temperature ($kT\sim 0.4$ keV) thermal emission and either 
power-law or high temeprature thermal emission.
It is unlikely that the whole soft X-ray emission is scattered light
of an AGN, because the optical reddeninig, \Av\ $\sim 3$, for the NLR
derived from the Balmer decriment (Koski 1978) predicts X-ray absorption
of at least \nH $= 6\times 10^{21}$\psqcm\ on the scattered continuum,
if the standard gas-to-dust ratio is assumed (and the gas is not ionized).
Since there is no evidence for such large absorption in the spectrum,
the soft X-ray emission is likely to originate from the region outside 
the nucleus, where presumably a starburst is taking place.
Results of the two-component fits to the soft X-ray data are shown
in Table 5. Here the Galactic absorption by \nH$= 1.1\times 10^{20}$\psqcm\
is assumed for the low temperature thermal spectrum. 
Whether the power-law or high temperature ($kT\geq 5$ keV) thermal spectrum
to be added to the $kT=0.45$ keV thermal model is difficult to distinguish 
at the present quality of data.

Excess emission is seen above 3 keV as well as an iron K line at
$\sim $6.4 keV (Fig. 3). 
The 3--10 keV continuum is very flat ($\Gamma = 0.1\pm 0.4$).
The flat spectrum is probably due to absorption of which the spectral
cut-off is masked by the soft X-ray component emitted outside the
X-ray absorbing material. If a power-law of $\Gamma = 1.8$ is assumed
for the spectrum of the absorbed source, the absorption column density 
is \nH $\simeq 4\times 10^{23}$\psqcm. An extrapolation of the power-law
is well below the upper limit given by the CGRO/OSSE observation
(Dermer et al 1997).
The 2--10 keV luminosity corrected for the absorption is 
$5\times 10^{42}$\ergps.
The iron K line feature is 
consistent with a single, narrow 6.4 keV line and there is no strong evidence 
for higher energy lines (see Fig. 3). The EW of the line is about 
$520^{+470}_{-260}$ eV.

An alternative model for the flat hard X-ray spectrum is reflection 
from cold material. A very large EW of iron line (e.g., $EW\geq $1 keV) is
often observed in the reflection spectrum. 
The reflection-dominated spectrum has a very hard continuum
above 10 keV and would have the 20--100 keV flux of 
$\sim 4\times 10^{-12}$\ergpspsqcm\ (estimated assuming a pure reflection case
of the XSPEC model, {\tt pexrav}, Magdziarz \& Zdziarski 1995), 
which is still consistent with
the OSSE upper limit.

\subsection{Arp220}

The double nucleus in Arp220 is clearly seen in near-IR images (Graham et al 
1990; Scoville et al 1998). 
The large far-IR excess in the energy distribution makes this galaxy 
one of the most luminous sources in the nearby universe within 100 Mpc.
The nuclear region shows a signature of high excitation gas
([OIII]$\lambda 5007$/H$\beta \sim 10$, Sanders et al 1988) while 
the emission-line nebula shows LINER characteristics at
larger radii (Heckman et al 1990).
This galaxy is a famous example of the ionization photon deficit problem
(DePoy et al 1987; Leitherer \& Heckman 1995), which casts a doubt on
the starburst interepretation for the major energy source in this galaxy
despite the lack of direct evidence for an AGN.


\begin{figure}
\vspace{-1cm}
\centerline{\psfig{figure=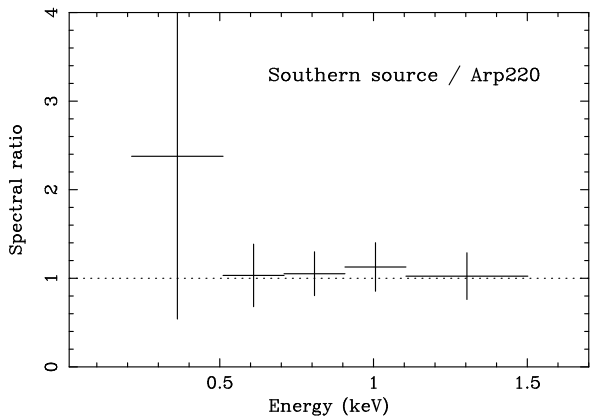,width=0.6\textwidth,angle=270}}
\caption{Plot of spectral ratio of the southern source and Arp220
made from the ROSAT PSPC data. The southern source appears to have
a similar spectrum and flux to Arp220 in the ASCA band ($\geq 0.5$ keV).}
\end{figure}


\begin{figure}
\centerline{\psfig{figure=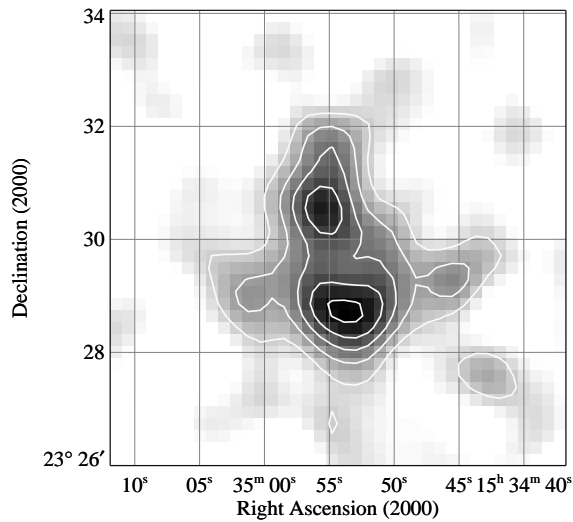,width=0.45\textwidth,angle=0}}
\centerline{\psfig{figure=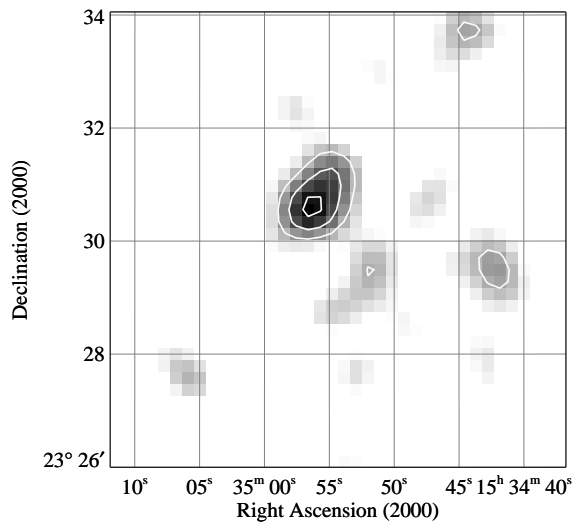,width=0.45\textwidth,angle=0}}
\caption{The ASCA SIS images of Arp220 region obtained in the energy bands of
0.5--2 keV (upper panel) and 2--5 keV (lower panel).
Arp220 and the southern source are resolved
in the soft band image whilst only Arp220 is detected in the hard band 
image.}
\end{figure}


\begin{figure}
\centerline{\psfig{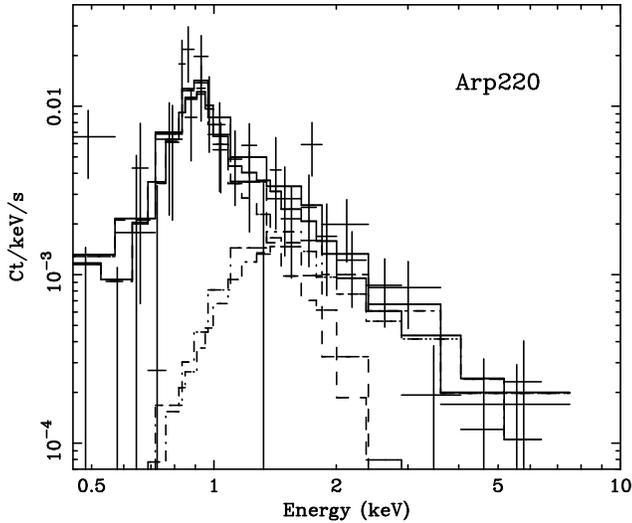}}
\caption{The ASCA SIS spectrum of Arp220. The data are fitted with a 
model consisting of a thermal emission spectrum and an absorbed power-law.
Note that the data are a sum of Arp220 and the southern source which 
contributes a half of the total flux below 2 keV with a similar
spectral shape to that of Arp220.}
\end{figure}

\subsubsection{The ASCA images and spectrum}

As the ROSAT PSPC image (observed between 1993 July 24 and August 3,
Heckman et al 1996) shows, another extended
X-ray source is observed to the south of Arp220 (``the southern source'').
Heckman et al (1996) associate the source with a distant ($z\sim 0.1$)
group of galaxies or a poor cluster. 
Optical spectroscopy of galaxies at the extended
X-ray source shows that those galaxies are at redshift of 0.08
(J. Hibbard, private communication) which supports the group/cluster 
interpretation.
No matther whether it is physically associated with Arp220,
both sources have similar soft X-ray spectra and fluxes 
(Fig 4; also see Table 1 in Heckman et al 1996).
Arp220 and the southern source are resolved also in the ASCA SIS 
0.5--2 keV image (Fig 5a).
However, in the energy band above 2 keV, no significant X-rays are
detected from the south source but only Arp220 (Fig. 5b).
As the energy spectrum shows (Fig 6), no X-ray emission is detected
above 5 keV even in the GIS image.

An image analysis estimates about half ($\sim$46 per cent) of 
the 0.5--2 keV observed counts is attributed to Arp220,
hence half the observed flux because of 
similar spectral shapes of the two sources in the energy band (
the Einstein IPC flux, e.g., David, Jones \& Forman 1992, 
was overestimated by a factor of $\sim $2).
At the spectral resolution of the SIS, an emission line peak due to
Fe L around 0.8--0.9 keV is evident, implying thermal emission.
There is also a faint hard X-ray component up to 5 keV.

The ASCA spectrum (Fig. 6) is modelled in a similar way to that for Mrk231.
Spectral fitting results are summarized in Table 4.
The X-rays from the `southern source' are contained in the ASCA spectrum. 
Observed fluxes in the 0.5--2 keV and 2--10 keV bands are
$1.7\times 10^{-13}$\ergpspsqcm ~and $2.2\times 10^{-13}$\ergpspsqcm,
respectively.
The characteristic temperature of the extended thermal emission is 
$kT = 0.76^{+0.13}_{-0.11}$ keV and the metallicity is 0.3($\geq 0.1$)\Zs.
This component has a 0.5--2 keV luminosity of $8.4\times 10^{40}$\ergps
~at the distance of Arp220 of which 
$\sim 3.9\times 10^{40}$\ergps ~comes from Arp220.
The hard component modelled by an absorbed power-law with a photon-index
of 1.8 requires absorption by \nH $= 6^{+12}_{-6}\times 10^{21}$\psqcm.
A photon index ranging between 1.2 and 4.8 is allowed for the power-law.
If a thermal emission model replaces the power-law,
its temperature would be about 2 keV but with large uncertainties.
The lack of significant X-ray detection at the iron K band means
no useful constraint available for an iron line strength.
The 2--10 keV flux of this hard component is $2.1 \times 10^{-13}$\ergpspsqcm
and a corresponding absorption-corrected luminoisty is 
$1.4\times 10^{41}$\ergps.

\subsection{NGC6240}


\begin{figure}
\centerline{\psfig{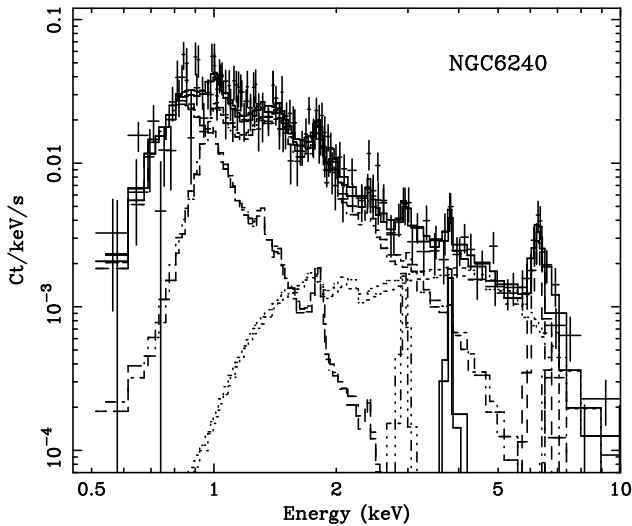}}
\caption{The ASCA SIS spectrum of NGC6240. The data are fitted with a 
model consisting of a two-temperature thermal emission spectrum and 
an absorbed power-law with emission-lines due to Ar, Ca, and Fe K$\alpha $.
Detailed spectral analysis is described in Iwasawa \& Comastri (1998).}
\end{figure}

Many properties of NGC6240 have been interpreted 
in terms of intensive star formation (e.g., Majewski et al 1993; 
Ridgway et al 1994; Shier, Rieke \& Rieke 1996). The near-IR spectrum is well
accounted for by a starburst stellar population. Extremely strong 
molecular hydrogen lines, e.g., H$_2$ S(1) at 2.1\micron, have been noticed 
(DePoy et al 1986; van der Werf et al 1993; Ridgway et al 1994).
The LINER-like optical spectrum 
and the large velocity field ($\sim 1000$ \kmps) in the nuclear 
region can be explained 
as results from an outflow driven by a starburst (Heckman et al 1990).
The extended H$\alpha$ nebula is $\sim 50\times 60$ kpc in size and
$\sim 10^{42}$\ergps\ in luminosity, which is larger and more powerful
than Arp220 despite the smaller infrared luminosity.
Luminous extended soft X-ray emission ($\sim 10^{42}$\ergps) 
has also been found with the ROSAT 
PSPC (Schulz et al 1998; Iwasawa \& Comastri 1998) and HRI (Komossa et al
1998). NGC6240 is the brightest X-ray source 
among the four FIRGs presented here.

The ASCA results on NGC6240 have been published by several authors
(Kii et al 1997; Turner et al 1997). A detailed spectral analysis
is reported in Iwasawa \& Comastri (1998).
Schultz et al (1998) and Netzer, Turner \& Geroge (1998) attribute
80 per cent and all the soft X-ray emission
to scattered AGN light, based on the ROSAT PSPC data and the ASCA data,
respectively.
However, the ASCA soft X-ray spectrum is modelled by two-temperature thermal
emission in Iwasawa \& Comastri (1998), who interpret it as
absorbed high-temperature gas and low-temperature, spatially more extended gas
produced by a starburst (see Fig 7). 
Since the HRI image shows that a large fraction ($>70$ per cent)
of the 0.1--2.4 keV flux is extended outside the PSF of the HRI  
($\sim $2.4 kpc in radius at NGC6240), 
we are in favour of the starburst interpretation
for the soft X-ray emission.
The central bright part of the soft X-ray nebula is nearly spherical
and has a size of about 10--20 kpc across (see Komossa et al 1998),
which may be related to the bright hourglass-shaped region ($10\times 7$ kpc)
of the H$\alpha$ nebula imaged by Heckman et al (1987).

The ASCA spectrum above 3 keV is very flat and shows a prominent
iron K line feature around 6.5 keV. 
The most likely interepretation of these spectral signatures in the 
hard X-ray band is reflection of a burried AGN radiation, predominantly from
cold, thick material. The lack of detection of an absorbed component
up to 10 keV implies that the column density of the matter occulting
any central strong X-ray source exceeds $\sim 2\times 10^{24}$\psqcm.
The similarity of the ASCA spectrum above 3 keV to NGC1068 (Ueno et al 1994;
Iwasawa et al 1997) renders an analogous argument on the 
luminosity of a central source possible, given a likely estimate of 
the central source of NGC1068 ($\sim 10^{44}$\ergps, Pier et al 1994).
The 3--10 keV luminosity of NGC6240, $2\times 10^{42}$\ergps, 
an order of magnitude above that of NGC1068, suggests that the FIRG 
may contain a QSO nucleus emitting at $\sim 10^{45}$\ergps.

\section{Composite X-ray spectrum}

\subsection{Thermal and nonthermal X-ray emission}

The X-ray spectra of FIRGs consists of multiple components 
in the 0.5--10 keV band. Thermal emission features are seen in the
soft X-ray band. Some fraction of observed soft X-ray emission may
be due to the scattered light of a hidden AGN (e.g., Mrk273).
However, as the ROSAT HRI images of Arp220 (Heckman et al 1996)
and NGC6240 (Komossa et al 1998) show, most of the soft X-ray emission
is extended, implying that thermal emission produced by a starburst
is the likely source. On the other hand, 
the hard X-ray spectrum is an important probe for AGN.
Particularly, the presence of a strong 6.4 keV line from Fe K$\alpha$
is a crucial diagnostic of obscured AGNs (Mrk273 and NGC6240).
The spectral shape and intensity of hard X-ray emission 
strongly depends on the absorption column density: the energy of the absorption
cut-off is a function of \nH, and when \nH\ exceeds 
$\sim 2\times 10^{24}$\psqcm, the absorbed direct emission from
a central source would disappear beyond the ASCA bandpass ($>$10 keV)
and leave the ASCA hard band spectrum dominated by weak reflected light
(NGC6240).

To characterize the X-ray spectra of the FIRGs, 
it is useful to compare with X-ray spectra showing characteristic 
features of each component observed in well-known 
starburst and Seyfert-2 galaxies.
We selected a typical
sample of starburst (M82), Compton-thick Seyfert 2 (NGC1068 and NGC4945),
and Compton-thin Seyfert 2 (NGC4507 and NGC4388) galaxies.
All the ASCA SIS spectra are shown in Fig. 8 as well as the sample FIRGs.
Note that the data for Arp220 have been corrected for the contribution 
from ``the southern source''.

\subsection{Comparison sample}




\begin{figure}
\centerline{\psfig{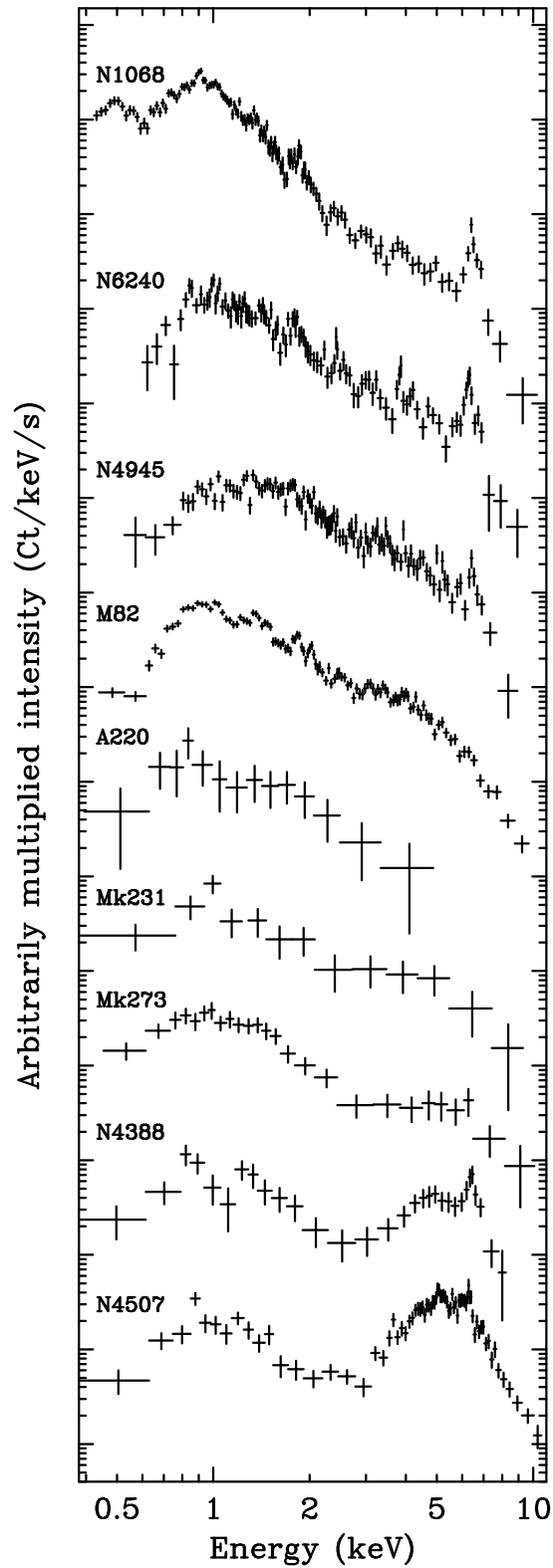}}
\caption{The ASCA SIS spectra of NGC1068, NGC6240, NGC4945, M82, Arp220,
Mrk231, Mrk273, NGC4388 and NGC4507 (from up to bottom). The y-axis is 
in unit of counts keV$^{-1}$s$^{-1}$ but arbitrarily multiplied 
for clarity. The Arp220 spectrum has been corrected for the 
contamination from `the southern source'. 
See details for individual objects
in Section 5.2 and for comparisons in Section 5.3.}
\end{figure}


\begin{table*}
\begin{center}
\caption{Ratios of the ROSAT PSPC count rate between the hard (0.5--2 keV)
and the full (0.1--2 keV) bands. The count rates are corrected for background.}
\begin{tabular}{rcccc}
 & Mrk231 & Mrk273 & Arp220 & NGC6240 \\[5pt]
0.5--2/0.1--2 keV & $0.62\pm 0.06$ & $0.73\pm 0.09$ & $0.92\pm 0.14$ &
$0.93\pm 0.11$ \\
\end{tabular}
\end{center}
\end{table*}

\subsubsection{M82}

M82 is a nearby ($D=3$ Mpc) well-studied starburst galaxy.
This galaxy shares many properties with the FIRGs presented in this paper
but less powerful in terms of infrared luminosity ($3\times 10^{10}$\Ls). 
Extended soft X-ray emission has been detected with the Einstein Observatory
(Fabbiano 1988). The large soft X-ray nebula extends along the superwind
structure, i.e., the minor axis of the edge-on galaxy disk.

The ASCA spectrum of M82 has been studied by several authors 
(Moran \& Lehnert 1997; Ptak et al 1997; Tsuru et al 1997).
Its multi-temperature nature has been known from spectral analysis
(see Tsuru et al 1997).
There is significant hard X-ray emission above 3 keV 
($\sim 2\times 10^{40}$\ergps), but the iron K feature is very faint.
Although the origin of the hard X-ray emission is not entirely 
understood, the presence of ArXVII(3.1 keV) and CaXIX(3.9 keV) suggests
that a significant fraction is of thermal origin. 
The BeppoSAX data 
rule out a power-law spectrum (M. Cappi, private com.). 
The significant detection of a Fe K line at 6.7 keV in the BeppoSAX 
spectrum of a similar starburst galaxy NGC253 strongly supports the
thermal emission interpretation (Mariani et al 1998).

The observed 0.5--10 keV luminosity is $3.2\times 10^{40}$\ergps.
Extended soft X-ray emission has been resolved with the ASCA SIS
because of the proximity of the galaxy. It has been found that 
that the X-ray spectrum is softer in the outer region than
in the central region.

\subsubsection{NGC1068}

NGC1068 is a classical Compton-thick Seyfert 2 galaxy ($D=14.4$ Mpc).
Since the discovery of polarized broad-line emission by Antonucci
\& Miller (1985) this galaxy provides the strongest case for the 
unification scheme of both types of Seyfert galaxies.
NGC1068 is also a powerful infrared source ($2\times 10^{11}$\Ls, Telesco 
et al 1984).
In the circumnuclear region, a dense molecular disk of a 100 pc scale
(Jackson et al 1993; Tacconi et al 1994) and kilo parsec scale star
forming arms (Helfer \& Blitz 1995) have been found.
A clear difference from the FIRGs is its clear view to the Seyfert 2 nucleus,
i.e., the NLR. The reddening to the NLR is small (\Av $\sim 0.2$).

In X-rays, the detection of a strong Fe K line with Ginga (Koyama et
al 1989) made this object the first example of Compton-thick Seyfert
2s.  No detection of an absorbed X-ray source up to 100 keV means that
the line-of-sight column density should be far above $10^{25}$\psqcm
(Matt et al 1997).  A sub-parsec scale, nearly edge-on (inclination of
larger than 80$^{\circ}$) disk has been discovered by high-resolution
radio imagings in H$_2$O maser and continuum (Greenhill et al 1996;
Gallimore et al 1997), which presumably hides a central X-ray source.

The soft X-ray emission is spatially
extended (Wilson et al 1992) and almost certainly associated with 
starburst activity in the circumnuclear region. The soft X-ray portion of
ASCA spectrum ($\leq 3$ keV) is dominated by thermal emission described by
a two-temperature model (Ueno et al 1994). The X-ray properties of NGC1068
is reviewed by Wilson \& Elvis (1997).
As shown in Fig. 8, the X-ray spectrum shows an
enormous excess in soft X-ray.

\subsubsection{NGC4945}

NGC4945 is a nearby ($D=3.9$ Mpc) FIRGs containing a 
{\it nearly} Compton-thick nucleus.
This galaxy is a far-IR excess galaxy with 
similar properties and luminosity to M82. 
The nucleus is heavily reddened and 
there is little evidence for an active nucleus apart from hard X-ray band. 
Superwind features, e.g., a hollow-cone shaped optical image (Nakai 1989),
split optical emission line profiles (Heckman et al 1990), have been
found. The optical emission-line gas in the nuclear region shows 
LINER type excitation (Whiteoak \& Gardner 1978).

A variable hard X-ray (10--30 keV) source was identified with this galaxy 
using Ginga observations, which is heavily absorbed by a column density of 
$\sim 5\times 10^{24}$\psqcm\ thus only visible above 10 keV 
(Iwasawa et al 1993). The detection of bright hard X-ray emission ($\sim$
100 keV) by the CGRO/OSSE confirmed it (Done et al 1996).
The absorption-corrected 2--10 keV luminosity is estimated to be
$2\times 10^{41}$\ergps.
Bright H$_2$O maser emission ($\sim 57$\Ls) has been found 
(Dos Santos \& Lepine 1979; Baan 1985; Braatz, Wilson \& Henkel 1996; 
Greenhill, Moran \& Herrinstein 1997). 
Like NGC1068, the water maser disk can be roughly
approximated to a Keplarian disk on parsec scale (Greenhill et al 1997).

A strong Fe K complex and reflection continuum are observed at the nuclear
position of the galaxy and an extended X-ray nebula with some point sources 
(e.g., Iwasawa 1995; Brandt, Iwasawa \& Reynolds 1996) are also resolved. 
Therefore, the X-ray emission observed with ASCA consists both of the reflected
AGN emission and starburst emission.
Since the bandpass of ASCA is limited to energies below 10 keV, the absorbed
component is invisible.
The absorption in the soft X-ray band ($<$ 1 keV) is 
mainly due to the large Galactic 
extinction (\nH\ $\sim 1.5\times 10^{21}$\psqcm, Dickey \& Lockman 1990)
because of the low Galactic latitude ($b\sim 15^{\circ}$).

\subsubsection{NGC4388 and NGC4507}

These two Seyfert 2 galaxies contain typical Compton-thin sources
absorbed by \nH $\approx (4-5)\times 10^{23}$\psqcm.
In the ASCA band, the the direct radiation from the central source is 
still visible above the absorption cut-off at 4--5 keV.
With this amount of absorption, a strong hard X-ray excess is seen,
often associated with a 6.4 keV Fe K line.
The heavly absorbed hard X-ray component is more prominent in NGC4507 
than NGC4388.
The ASCA results have been reported by Iwasawa et al (1997) for NGC4388 and 
Comastri et al (1997) for NGC4507.
The estimated 2--10 keV luminosities corrected for the absorption
are $2\times 10^{42}$\ergps\ for NGC4388 and
$2\times 10^{43}$\ergps\ for NGC4507.

The weak soft X-ray emission is mainly due to the scattered light 
in NGC4507 (Comastri et al 1997) while a significant fraction of 
the soft X-rays observed below 2 keV in NGC4388 orginates in the extended 
thermal emission resolved with the ROSAT HRI (Matt et al 1994).

\subsection{The X-ray spectra of the FIRGs and the comparison sample}

In Fig. 8, almost entire or a significant fraction of soft X-ray
emission is due to thermal emission originating from starbursts in all
galaxies except for NGC4507.  Note that the exceptionally large soft
X-ray excess emission in NGC1068 compared with the other
galaxies, which may be related to absorption in the starburst regions,
as discussed in Section 6.1.2.  The spectrum of M82 is presumably of a
typical starburst without any significant AGN component. A clear
difference from the other Seyfert 2 spectra is the absence of a strong
Fe K line.

The two Compton-thick Seyfert 2s (NGC1068 and NGC4945) show very
strong Fe K features ($EW>1$ keV), indicative of reflection-dominated
spectrum.  The detection of a similar Fe K feature in NGC6240 has
revealed a Compton-thick source in the FIRG despite the rest of the
ASCA spectrum resembling M82.

The two Compton-thin Seyfert 2 galaxies (NGC4388 and NGC4507) show
clear hard X-ray excess due to strong absorption.  The 6.4 keV Fe K
lines with {\it EW} of few hundreds eV are also seen.  Mrk273 is most
likely classified as this type but its hard X-ray excess is less
prominent.  The relatively flat 3--10 keV spectrum suggests that
Mrk231 is also the case. As shown in the spectral analysis (Section
4.1.2 and Fig. 1) the absorption would be smaller (few times of
$10^{22}$\psqcm) than the other three objects. However, the origin of
the hard X-ray emission is not yet clear from the present data and several
possibilities are discussed in Section 6.2.2.  For the Compton-thin
sources, sources with a stronger hard X-ray excess are placed lower in
Fig. 8.

Arp220 lacks hard X-ray emission above 5 keV and the ASCA spectrum 
apears to be softer than the other objects, although the signal to
noise ratio of the data is poor.

\subsection{The ROSAT band colour}

The ROSAT bandpass extends down to 0.1 keV, which is more sensitive to
the low temperature thermal emission than ASCA.
The hardness ratio obtained from the ROSAT PSPC data for the four FIRGs
are shown in Table 6, to assess their soft X-ray band (0.1--0.5 keV)
spectra. 

Arp220 and NGC6240 have harder spectra than those of Mrk231 and Mrk273
in the ROSAT band. It is uncertain whether the hardness is due to
a higher averaged temperature or larger absorption.

\section{discussion}


\begin{table*}
\begin{center}
\caption{Far-infrared, 2.2\micron, optical emission-line (H$\alpha$+[NII]), 
and X-ray luminosities (erg\thinspace s$^{-1}$) of the FIRGs and M82. 
$L$(FIR) and $L$(H$\alpha$)
are taken from Armus, Heckman \& Miley 1990. Note that values listed
in Lonsdale et al (1985) are used for $L$(FIR). The 2.2\micron\ power is 
calculated using the measurements by Ridgway et al (1994).
The $L$(H$\alpha$) is 
total luminosity for the extended emission line nebula (no correction for
reddening). L(SX) and L(HX) are observed X-ray luminosities 
in the 0.5--2 keV and 2--10 keV bands, 
respectively (no correction for absorption). }
\begin{tabular}{lccccc}
Galaxy & log L(FIR) & log P(2.2\micron ) & log L(H$\alpha$) & log L(SX) & 
log L(HX) \\
& \ergps & erg\thinspace s$^{-1}$\micron $^{-1}$ & \ergps & \ergps & \ergps
\\[5pt]
Mrk231 & 45.69 & 45.20 & 42.39 & 41.65 & 42.32 \\
Mrk273 & 45.45 & 44.23 & 42.21 & 41.64 & 42.19 \\
Arp220 & 45.49 & 43.52 & 41.04 & 40.61 & 41.06 \\
NGC6240 & 45.11 & 44.38 & 42.26 & 41.84 & 42.35 \\[5pt]
M82 & 44.50 & 42.15 & 41.10 & 39.99 & 40.35 \\
\end{tabular}
\end{center}
\end{table*}

\subsection{Starburst properties}

\subsubsection{X-ray luminosities}

Based on the spectral analysis of the ASCA data, we assume that the observed
soft X-ray emission below 2 keV is predominantly thermal emission 
originating from a starburst in those galaxies.
Then we compare the soft X-ray luminosities with the other possible indicators
of a starburst. We note that it is premature to claim any correlation from
these comparisons because of the small size of the sample, and 
the trends seen here must be tested by a larger sample.

Far-IR luminosity is generally thought to
be a good indicator of starburst intensity.
However, Arp220 is far more powerful in the far-IR, relative to other
bands, e.g., the soft X-ray emission (Fig. 9), H$\alpha$, optical
light ($L_{\rm B}$, see Table 1), than the other FIRGs.
The far-IR excess in Arp220 is discussed in Section 6.3

Conceivably, starburst-driven galactic outflow is a main heating source
of the extended optical emission-line nebulae (Heckman et al 1990). 
Armus et al (1990) show that more than half of the total 
H$\alpha$+[NII] luminosity (not corrected for internal reddening)
comes from outside the central $2\times 2$ kpc box and the sizes of the
nebulae are a few tens kpc in the FIRGs apart from Mrk231,
in which the broad H$\alpha$ line observed in the nucleus dominates the 
total luminosity.
Shock heating of the ISM by the outflow is also a likely source of the 
soft X-ray emission, as supported by the soft X-ray morphology aligned with 
a superwind structure in nearby starburst galaxies like M82 (e.g., Fabbiano 
1988; Strickland et al 1997).
In fact, the luminosity ratio of the H$\alpha$+[NII] and 
soft X-ray (0.5--2 keV) of the FIRGs are similar to each other (2.6--5.5;
Fig. 10 ) while M82 has a larger value ($\sim 13$).

On the other hand, a starburst stellar population appears to account for
the near-IR emission in the FIRGs except for Mrk231 (Ridgway et al
1994; Shier et al 1996). A correlation between the 2.2\micron\ and the
soft X-ray luminosity is relatively good (Fig. 11).
The excess of the 2.2\micron\ emission in Mrk231 is probably due to 
a contribution from the AGN, as suggested from the near-IR properties
(Fig. 12; blue $H-K$ colour and the very shallow CO absorption feature
at 2.3\micron).
It should be noted that optical depths at 2.2\micron\ and the soft X-ray 
band are similar to each other if the standard gas-to-dust ratio is 
assumed. 


\begin{figure}
\centerline{\psfig{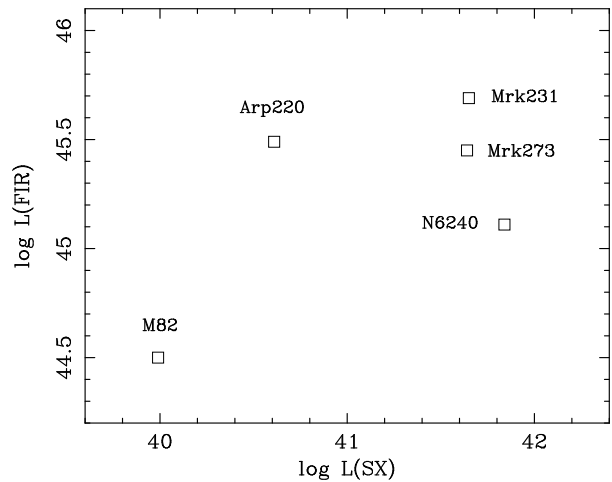}}
\caption{Plot of the far-IR luminosity against 
the 0.5--2 keV (SX) luminosity (Table 7). Note the large excess in 
the far-IR of Arp220.}
\end{figure}


\begin{figure}
\centerline{\psfig{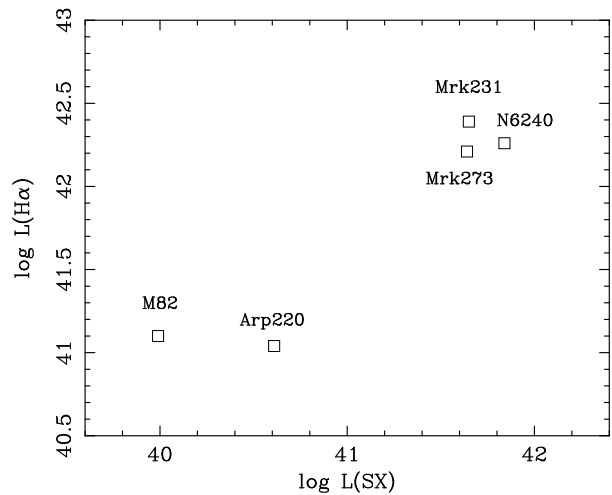}}
\caption{Plot of the total H$\alpha $+N[II] luminosity against 
the 0.5--2 keV (SX) luminosity (Table 7). }
\end{figure}


\begin{figure}
\centerline{\psfig{figure=sx-k.ps,width=0.45\textwidth,angle=270}}
\caption{Plot of 2.2\micron\ power (\ergps\micron$^{-1}$) against 
the 0.5--2 keV (SX) luminosity (Table 7). }
\end{figure}


\begin{figure}
\centerline{\psfig{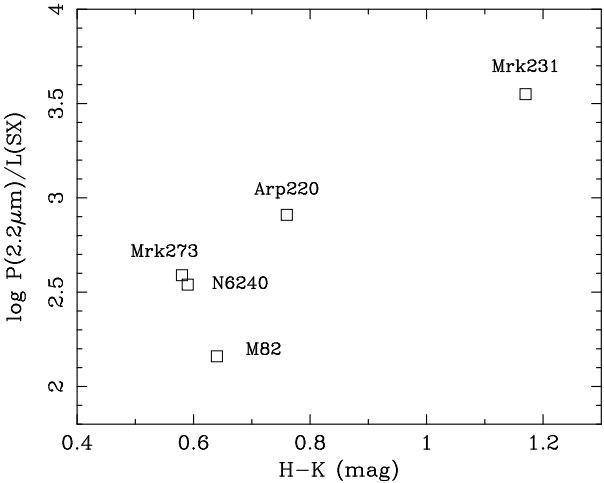}}
\caption{Plot of the luminosity ratio between 2.2\micron\ and the
soft X-ray (0.5--2 keV) emisison against the near-IR colour ($H-K$).
The values of $H-K$ are taken from Ridgway et al (1994 and references 
therein).}
\end{figure}

In summary, the soft X-ray emission, extended optical emission-line nebulae,
and near-IR emission are consistent with each other in the 
context of starburst, but not with the far-IR emission.

\subsubsection{Temperature and absorption structure in the X-ray nebulae}

At the low quality of the present data, a single temperature model is
a sufficient description for the starburst soft X-ray emission.
However, some good quality data like of M82 (Tsuru et al 1997;
Strickland et al 1997) and NGC6240 (Iwasawa \& Comastri 1998)
have provided evidence for multi-temperature gas and absorption
for the starburst regions.
Obscured, hot gas in the near-nuclear region and cooler gas at
larger radii are expected from the superwind scenario (Chevarier \&
Clegg 1985; Tomisaka \& Ikeuchi 1988).
High obscuration in starbust nuclei have also been inferred from optical
and infrared observations of starburst (e.g., M82, Heckman et al 1990)
and ultraluminous infrared galaxies (Roche et al 1983;
Thronson et al 1990; Lutz et al 1996; Dudley \& Wynn-Williams 1997).

The size of the starburst region in far-IR galaxies appears to be
compact (dense molecular disk in Mrk231 and Arp220 are inferred 
to be few hundreds pc in radius, Bryant \& Scoville 1996; 
Scoville et al 1997). A compact starburst near the nucleus has also
been found in some Seyfert 2 galaxies (e.g., Mrk477, Heckman et al
1997). In contrast, the starburst ring in NGC1068 is on rather large
scale (few kpc) and well separated from the Seyfert 2 nucleus 
(Neff et al 1994).
The compact starburst regions are likely to be more obscured than
the large scale starburst region in NGC1068. 
The much softer spectrum of NGC1068 in the low energy X-ray band
than the other FIRGs (Fig. 8) is therefore suspected to be due 
to the difference of obscuration.
The correction for absorption would be important for 
estimating the true starburt luminosity through X-ray data.
High quality data from future observations are required to decompose 
the gas structure.

\subsection{AGN properties (Mrk273, NGC6240 and Mrk231)}

\subsubsection{Mrk273 and NGC6240}


The absorption-corrected 2--10 keV luminosity of Mrk273 is 
only $5\times 10^{42}$\ergps, nearly three orders of magnitude below
the far-IR luminosity. This suggests that Mrk273 contains an obscured AGN
which is not a QSO nucleus which would provide the bulk of 
the bolometric luminosity, but a moderate-luminosity Seyfert.
However, note that since a reflection-dominated spectrum 
is not ruled out by the 
present data (see Section 4.2.2), the intrinsic luminosity of the hidden AGN 
could be as high as QSOs.

The heavily obscured X-ray nucleus in the nearby Seyfert 2 galaxy NGC4945
has the 2--10 keV luminosity of $\sim 10^{41}$\ergps\ after
correction for absorption. It
is also three orders of magnitude below the far-IR luminosity.
In these cases, the obscured AGN do not play a major role in powering
the far-IR emission and starbursts are probably the most
important energy source. In fact, the mid-IR properties of NGC4945 
are in excellent agreement with those of the other starburst galaxies (Genzel
et al 1998). 
The lack of AGN features in the mid-IR band can easily be understood,
considering the extremely thick obscuration of the central source in
NGC4945 implied 
from X-ray observations (\nH $\simeq 5\times 10^{24}$\psqcm, or 
$\tau_{30\mu {\rm m} }\sim 25$, Iwasawa et al 1993; Done et al 1996).
The detection of high excitation mid-IR lines,
e.g., [OIV] 26\micron, [NeV] 14.3,24.3\micron, 
in Mrk273 from the ISO/SWS (Genzel et al 1998)
is probably due to the smaller obscuration of the central source
(\nH $\sim 5\times 10^{23}$\psqcm, $\tau_{30\mu {\rm m}}\sim 2$). 
Alternatively, perhaps the X-ray source has now faded to its low state,
leaving the large EW ($\sim 500$ eV)
of the Fe K line from a distant 
line emitter as a reverberation effect
(e.g., Mrk3, Iwasawa et al 1994; NGC2992, Weaver et al 1996).

NGC6240 is suspected of harbouring
a QSO nucleus, possibly emitting at $10^{45}$\ergps. 
The estimated QSO luminosity, almost comparable with the far-IR luminosity,
suggests that the hidden QSO makes a significant contribution to the far-IR 
(bolometric) luminosity and may even dominate it.
The galaxy is the most powerful soft X-ray emitter among 
the four FIRGs presented here, indicating that the starburst is also
powerful.
Having both powerful an AGN and a starburst as energy sources, despite the 
small far-IR luminosity (Table 7) compared to the other FIRGs 
may be a problem.
If the superwind heating the soft X-ray/H$\alpha$ nebula is mainly driven by a
QSO instead of starburst, this problem could be relaxed.

The estimate of the intrinsic AGN luminosity in NGC6240 relies on 
the bolometric (optical to X-ray) luminosity of a hidden Seyfert 1 nucleus 
in NGC1068 ($3\times 10^{44}$(14 Mpc/$D)^2(0.01/f_{\rm refl})^{-1}$\ergps,
where $D$ is the distance to NGC1068 and $f_{\rm refl}$ is the fraction of 
light from the hidden nucleus reflected into our line of sight) 
given by Pier et al (1994). 
The validity of their estimate is justified by the fact that they used
polarization measurements, which are expected to be affected little by
possible stellar light contamination, otherwise the derived intrinsic
luminosity of NGC1068 (and thus NGC6240) would be overestimated.

The central source of NGC6240 should be occulted by thick material with
a column density over $2\times 10^{24}$\psqcm\ along the line of sight.
Since no strong evidence for an AGN has been found in this galaxy,
apart from the hard X-ray spectrum, the powerful starburst 
masks the AGN phenomenon in a more heavily-obscured region inside.

\subsubsection{Mrk231}

Although the X-ray emission with the hard spectrum above 3 keV 
may be attributed to an AGN, the observed X-ray luminosity is quite
small ($L_{\rm 2-10keV}\sim 2\times 10^{42}$\ergps), compared to the
infrared luminosity and the reddening-corrected optical luminosity
($M_{\rm V}\sim -25.1$, Boksenberg et al 1977), both of which are 
in the range of QSOs. 

The nucleus of Mrk231 clearly shows manifestations of a Seyfert-1
or BAL QSO type active nucleus in various wavebands.
The broad (FWHM$\sim 4200$ \kmps) H$\alpha$ was measured at 
luminosity of $7\times 10^{42}$\ergps\ by Boksenberg et al (1977),
while Armus et al (1990) measured $2\times 10^{42}$\ergps\ within
the central $2\times 2$ kpc region.
The observed small X-ray luminosity is difficult to understand
as a normal Seyfert 1 or QSO. 
The good correlation between 2--10 keV and H$\alpha$ luminosity 
(log[L(2--10keV)/L(H$\alpha $)]$\sim 1$) found in Seyfert 1 galaxies
and QSOs (Ward et al 1988) predicts that the 2--10 keV luminosity 
of Mrk231 would be (2--7)$\times 10^{43}$\ergps, which is at least
10 times higher than the observed value.

The difference between the two H$\alpha$ measurements could be 
due to variability of the line, and may suggest a declining
central source. An inspection of the X-ray data from earlier mission like
HEAO-1 however suggests that the past activity in the hard X-ray band 
was also weak ($2\times 10^{-11}$\ergpspsqcm; the X-ray flux expected 
from the (2--10keV)--H$\alpha$
correlation and the Boksenberg et al (1977) measurement is below 
the flux limit ($3.1\times 10^{-11}$\ergpspsqcm\
from the HEAO-1 A2 survey by Piccinoti et al 1982; Rieke 1988
gave an upper limit of $6.5\times 10^{-12}$\ergpspsqcm\ based on the HEAO-1
A1 survery).
Possible explanations for the X-ray quietness are
discussed below.

\leftline{\it (i) A powerful starburst (absence of AGN)}

\ni The hard X-ray component may be attributed to high temperature 
gas produced by a starburst. A powerful starburst which is two orders of 
magnitude more luminous than M82 is required if simply scaled.
The lack of detection of of high excitation lines
in the ISO/SWS mid-IR spectrum of Mrk231 
and inferred large reddening (Rigopoulou et al 1998) 
may support this hypothesis.
However, many properties, not usually seen in starburst galaxies but in
AGNs (e.g., broad H$\alpha$, strong optical FeII emission, etc.),
have to be explained by a starburst alone. 
The variability in one of the absorption-line systems (Boroson 
et al 1991; Kollatschny, Dietrich \& Hagen 1992) and nonthermal radio 
continuum (Condon et al 1991) 
does not favour this hypothesis but a compact
single object, i.e., AGN. 
The dominance of a nonstellar source in the near-IR band is also suggested
from the very weak CO absorption band 
(Majewski et al 1993; Ridgway et al 1994;  Shier et al 1996).
Therefore there are many difficulties to explain without 
a powerful AGN.

\leftline{\it (ii) Small-scale occultation and reflection of a central source}

\ni The weak X-ray luminosity could be explained by a reflection-dominated 
object. A similar 2--10 keV X-ray luminosity to that of 
NGC6240 implies that the intrinsic luminosity of the hidden central
source would be as large as $10^{45}$\ergps. 
This hypothesis is, however, inconclusive without detection 
of a strong Fe K line, and the apparent Seyfert 1 nature needs explanation.
The loose upper limit on the EW of the Fe K line ($< 1$ keV)
cannot rule out this possibility.

In this hypothesis, the central source must be absorbed completely
even in the X-ray band. However, the reddening implied for 
the UV/optical continuum (\Av$\sim 2$, Boksenberg et al 1977;
Hamilton \& Keel 1987; Smith et al 1995) is far too small.
To hide only a central X-ray source behind the optically thick matter, 
the obscuration must occur
on very small scale (well inside the broad line region).
The variability time scale (a few years) of the absorption system 
(Boroson et al 1991) suggests that the inner 1pc of the nucleus is
seen. Since X-ray emission
appears to be generated within 10 Schwarzschild radii ($< 10^{-4}$ pc for
a $10^8$\Ms\ black hole), the occultation has to occur on scales between
$10^{-4}$ pc to 1 pc. Disk warping (Maloney, Begelman \& Pringle 1996) 
may be responsible for this, although how a large column density in the 
line of sight can be achieved is uncertain.

\leftline{\it (iii) An X-ray quiet central source}

\ni Assuming that we are seeing 
an intrinsically X-ray quiet QSO nucleus through moderate
obscuration by low ionization gas, may be a straightforward explanation.
BAL QSOs have been recognized as significantly weak soft X-ray sources
from ROSAT observations (Green \& Mathur 1996).
Whether it is due to large absorption or low luminosity intrinsic
to the central sources is still uncertain (Mushotzky 1997),
although there is one example showing that absorption seems to be
the case (Mathur, Elvis \& Singh 1995).
If the hard X-ray emission detected with ASCA is the direct radiation 
from the nucleus, the X-ray source in Mrk231 would be intrinsically
X-ray quiet ($L_{\rm 2-10keV}\simeq 2\times 10^{42}$\ergps)
as well as being absorbed (\nH $\sim 2\times 10^{22}$\psqcm, see Table 4). 

This X-ray quiet hypothesis would also bring a serious problem with
the X-ray heating (photoionization) model of FeII (e.g., 
Netzer 1987; Collin-Souffrin et al 1988).
The intrinsic deficit of hard X-ray photons fails to explain the fact
that Mrk231 is one of the `extreme FeII emitters' (FeII/H$\beta\approx 2.1$,
Lipari 1994). The correlation between soft X-ray slope and EW(FeII) found 
by Wilkes, Elvis \& McHardy (1987) 
using the data from the Einstein Observatory 
is also in the opposite sense to the standard photoionization model.

In order to provide sufficient ionizing UV photons to account for
the observed large luminosities of Balmer lines and the optical--UV 
continuum, a large excess in the UV--soft X-ray band, which would
however be unobservable because of the moderate absorption, 
is required. This is possibly related to 
the class of ultrasoft AGNs (or narrow-line Seyfet 1s, Boller, Brandt \&
Fink 1996) which, too, often show strong FeII emission.

Alternatively, the bulk of the energy for the optical/UV emission
may be supplied from a starburst and mechanical heating by the
starburst wind is responsible for the FeII excitation,
as proposed by Lipari, Colina \& Macchetto (1994) and Lipari, Terlevich
\& Macchetto (1993).

\subsection{The large far-IR excess in Arp220}

The energy distribution of the Arp220 is characterized by the enormous
far-IR excess, which places the galaxy at the early stage of the
evolution scenario of infrared galaxies proposed by Sanders et al
(1988).  The lack of direct evidence for a QSO-like central source
favours a powerful, young starburst as the major source of the
bolometric luminosity.  A handful of hints of AGN activity, the high
[OIII]/H$\beta$ ratio in the nucleus (Sanders et al 1988) and a
compact radio core (Lonsdale, Smith \& Lonsdale 1995) can be
reconciled with photoionization by hot stars (e.g., `warmer',
Terlevich \& Melnick 1985) and a cluster of radio supernovae (Smith,
Lonsdale \& Lonsdale 1998), respectively.  A highly obscured starburst
can account for the `ionizing photon deficit' problem (DePoy et al
1987; Leitherer \& Heckman 1995; Armus et al 1995). 

No detection of strong hard X-ray emission ($>5$ keV) fails to provide
any evidence for a powerful AGN.  However, this does not readily rule
out the possibility of a deeply buried AGN as a major source of the
far-IR luminosity, because the starburst in Arp220 does not appear to
be sufficiently powerful, as discussed below.

The observed H$\alpha$ and soft X-ray emission in Arp220 are
consistent with the superwind scenario (Heckman et al 1996; Section
6.1.1).  Importantly, in the superwind model, the luminosity of the
nebular emission depends on the mechanical luminosity of starburst
winds.  Therefore the soft X-ray and H$\alpha$ emission in Arp220, one
order of magnitude less luminous than NGC6240, may suggest that the
starburst taking place in Arp220 is much weaker than in NGC6240
despite the larger far-IR luminosity.  This speculation is consistent
with the conclusion of Scoville et al (1997), based on the weak
mm-waveband free-free emission where dust is no longer optically
thick.

A possible explanation for the weak mechanical luminosity can be found
if an upper mass cut-off (say, at 30\Ms) in the initial mass function
(IMF) is assumed.  This argument has also been discussed for the
ionizing photon deficit problem.  The absence of a population of
massive stars would result in a smaller kinetic energy of the
starburst wind which is provided predominantly by Type II supernovae
and Wolf-Rayet (WR) stars while low mass stars power the far-IR
luminosity.  Few WR stars, that evolve from O stars with an initial
mass of larger than 20--40 \Ms, would exist.  The lack of WR feature
(at $\lambda\sim 4660$\AA) in the optical spectrum is consistent with
this.  Recombination lines from HII regions would also remain weak
because of fewer ionizing stars (stars with $M\geq $20\Ms).  The
rather `tepid' IRAS colour agrees with the dominance of low mass stars
as a primary heating source of the far-IR emission, since dust grains
heated by such population of stars mainly contribute to the IRAS
60\micron\ and 100\micron\ range (Mouri \& Taniguchi 1992).

Although this idea of a truncated IMF may partly solve the problem, 
the star formation rate estimated from the far-IR luminosity 
still overproduces the mechanical luminosity by a factor of
$\sim$5 more than needed to account for the observed soft 
X-ray/H$\alpha$ luminosities (Heckman et al 1996).
We therefore conclude that the starburst in Arp220 is less 
powerful than the other FIRGs in terms of the mechanical luminosity.

A deeply buried starburst (\Av$\sim 50$, Sturm et al 1996) inferred from 
the ISO spectroscopy is not seen in the X-ray data.
The column density \nH $\sim 1\times 10^{23}$\psqcm, deduced from
\Av$\sim 50$ is no longer opaque in the hard X-ray band (e.g., $> 3$ keV).
However, no hard X-ray emision expected from such an obscured starburst 
is detected in the ASCA spectrum.
The observed 2--10 keV luminosity of Arp220 (Table 7) is 
a factor of 6 smaller than that scaled from the luminosity
of M82 using the far-IR luminosity.
At least the X-ray data do not verify a starburst in 
a heavy (\Av$\sim 50$) obscuration to be particularly powerful.

If the far-IR emission of Arp220 is instead powered by an AGN, it must be a
Compton-thick source.  Because of their faintness in X-ray,
Compton-thick AGNs are usually hard to detect. Only nearby objects like
NGC1068 were detectable in the past X-ray surveys.  However, such
sources appear to be common.  A recent BeppoSAX survey has found a
number of Compton-thick Seyfert 2s (Salvati et al 1997).  The other
far-infrared excess galaxies such as NGC4945 (Iwasawa et al 1993; Done
et al 1996), the Circinus galaxy (Matt et al 1996) and 
NGC6240 (Iwasawa \& Comastri 1998) also contain AGNs
hidden behind obscuration with \nH$> 10^{24}$\psqcm.  A difference
between these Compton-thick sources and Arp220 would be the covering
factor of the obscuration.  No detection of even the
reflection-dominated X-ray spectrum means a central X-ray source to be
covered by an extremely thick matter almost completely.
Very little hot dust irradiated by a central AGN would then be visible,
which is consistent with the cool characteristic temperature of the 
far-IR emission in the IRAS band.
Therefore a starburst taking place at the smaller optical depth 
($\tau_{\rm dust}\sim 1$, or \Av$\sim 50$) 
would dominate the spectrum from near-IR to mid-IR as observed.

\subsection{Implications for other ultraluminous galaxies}

The far-IR dominance in the spectral energy distribution is similar between
FIRGs and Seyfert 2 galaxies.
There are indications that starbursts can be a significant energy source
in Seyfert 2 galaxies (Heckman et al 1995, 1997) and the outer parts of
the obscuring matter of AGNs may be the place where the starbursts are 
occurring.
Perhaps powerful FIRGs are even more heavily obscured version of Seyfert 2s.
Starbursts appear to dominate the emission of FIRGs 
in the near-IR to mid-IR band.
However, the optical depth to the starburst regions is still small
compared with that to the hidden X-ray sources in some FIRGs 
(NGC6240, NGC4945).
Arp220 might also be the case.
When the obscuration of a central source is not thick enough (say, 
a few times of $10^{23}$\psqcm\ in \nH), AGN features would emerge in
the optical--mid-IR spectrum (Mrk273).

Some ultraluminous IR galaxies (ULIRGs) with a clear Seyfert-2 optical
spectrum or polarized broad-line region have been found to harbour
an absorbed X-ray source with \nH\ values of order of $10^{23}$\psqcm\
(Mrk273, this work; Mrk463E, Ueno et al 1996; IRAS20460+1925, 
Ogasaka et al 1997; IRAS23060+0505, Brandt et al 1997).
They can be considered as luminous versions of classical Seyfert 2 galaxies.
However, no significant X-ray detection has been reported for
hyperluminous galaxies at higher redshift, 
like IRAS F15307+3252 (Fabian et al 1996; 
Ogasaka et al 1997) and IRAS F10214+4724 (no detection with 80 ks ASCA
observations while 2$\sigma $ detection with ROSAT reported by Lawrence et al 
1994) despite clear Seyfert-2 characteristics of their optical emission-line
spectra (Cutri et al 1994; Rowan-Robinson et al 1991; Serjeant et al 1998). 
No detection, at least in the ROSAT band, has been reported for 
a further four hyperluminous galaxies,
IRAS F00235+1024, F12514+1027, F14481+4454 and F1537+1950 (Wilman et al 1998).
For IRAS P09104+4109 (Kleinmann et al 1988), cluster emission 
surrounding the hyperluminous
galaxy dominates the observed X-rays ($2\times 10^{45}$\ergps,
Fabian \& Crawford 1995).

The lack of strong hard X-ray emission in hyperluminous galaxies 
poses the same problem as with Arp220.
If a powerful AGN exists, only faint scattered light would be emitted
and the upper limit of the 
L(2--10keV)/L(FIR) ratio for IRAS F15307+3252 ($4\times 10^{-3}$,
Ogasaka et al 1997) is still consistent with the known cases of 
NGC1068 and NGC6240 for which the ratio is $\sim 10^{-3}$.
If an AGN is intrinsically weak, 
a central massive black hole may not have grown
enough to emit QSO luminosity while a powerful starburst provide the
bulk of the bolometric luminosity, which would explain the X-ray
quiet nature.

Fabian et al (1998) has proposed a nearly spherical obscuring geometry
on 100 pc scale rather than a pc scale toroidal one 
for the major contributors to the X-ray background (XRB).
ULIRGs may be extreme cases of much larger column density.
Primary obscuration of a central X-ray source 
(e.g., \nH $\gg 10^{24}$\psqcm) perhaps occurs 
at small radii ($\leq 1$ pc) otherwise total mass becomes exceedingly large
while molecular gas clouds, composing a moderate obscuration 
on hundreds pc scale, absorb the radiation escaping from the inner region.
The small opening fraction of the large-scale absorbing shroud makes 
the scattered light weak.
It should be noted that a high covering fraction of obscuration is
demanded to explain the XRB with the obscured AGN models (Setti \& Woltjer
1989; Madau, Ghisellini \& Fabian 1994; Celotti et al 1995; 
Comastri et al 1995). 

\section*{acknowledgements}

KI thank all the member of the ASCA team who operate the satellite
and maintain the software and database. A.C. Fabian, Y. Taniguchi and 
the referee are thanked for helpful comments. J. Hibbard is thanked for
information on his unpublished observation.
The ASCA observations were made by following PIs:
T. Nakagawa (Mrk273 and NGC6240), L. Armus (Mrk231) and 
the ASCA team (Arp220).
This research has made use of data obtained through the High Energy
Astrophysics Science Archive Research Center (HEASARC), provided by
NASA's Goddard Space Flight Center.
KI thank PPARC for support.

\end{document}